\documentclass[aip,jcp,reprint,longbibliography,superscriptaddress,final]{revtex4-1}
\usepackage{graphicx}
\usepackage{dcolumn}
\usepackage{bm}

\usepackage{showkeys}
\usepackage[usenames,dvipsnames]{color} 
\usepackage{mathrsfs}
\usepackage{amsmath}
\usepackage{amssymb}

\newcommand{\beq}{\begin{eqnarray}}
\newcommand{\eeq}{\end{eqnarray}}
\newcommand{\dd}{{ \! \! \rm d}}

\newcommand{\cw}{1}

\begin{document}

\title{The influence of van der Waals forces on droplet morphological transitions and solvation forces in nanochannels}

  \author{F. Dutka}
    \affiliation{Institute of Physical Chemistry, Polish Academy of Sciences, ul. Kasprzaka 44/52, 01-224 Warszawa, Poland}
  \author{M. Napi\'orkowski}
    \affiliation{Institute of Theoretical Physics, Faculty of Physics, University of Warsaw, ul. Ho\.za 69, 00-681 Warszawa, Poland}
  
\date{\today}

\begin{abstract}
The morphological phase transition between a sessile and lenticular shapes of a droplet placed in a nanochannel is observed upon increasing the droplet volume. The phase diagram for this system is discussed within the  macro- and mesoscopic approaches. On the mesoscopic level, the van der Waals forces are taken into account via the effective interface potential acting between the channel walls and the droplet. We discuss the contact angle dependence on the droplet volume and the distance between the walls; this angle turns out to be  smaller than the  macroscopic Young's angle. The droplet presence induces the solvation force acting between the channel walls. It can be both attractive and repulsive, depending on the width of the channel. 
\end{abstract}

\pacs{47.60.Dx,68.03.Cd,68.05.-n,68.65.-k,83.50.Ha}    
                       
\keywords{nanofluidics, phase transitions, van der Waals forces, line tension, solvation forces}

\maketitle

\section{Introduction}
The progress in miniaturization of microfluidic systems brings new challenges for the theoretical description of such systems. The behavior and manipulation of liquid droplets or gas bubbles (called the discrete phase) in a planar channel of micrometer size  filled with immiscible continuous phase is rather well understood \cite{Squires2005,Seemann2012}. In the absence of electrostatic interactions and neglecting the gravity (which plays  minor role on these scale) the droplet can be described by the macroscopic theory \cite{Gennes2004,Bonn2009}. 

When the size of the channel becomes smaller and its height is below 100 nm, the droplet shape cannot be described by the macroscopic theory; one has to take into account the long-ranged van der Waals forces. They give rise to the effective interaction between the walls of the channel and the droplet surface \cite{Wong1992,Wong1992b}. In our mesoscopic description we consider droplets which do not touch the walls of the channel, and a thin layer of continuous phase separating the walls and the droplet is present \cite{Ajaev2012}. This type of morphologies  will be investigated in the following analysis. Fabrication of nanochannels and filling them with liquid is already experimentally feasible \cite{Kim2009,Shui2011}; carbon nanotubes are good examples of such nanocapillaries \cite{Megaridis2002,Mattia2008}. 

The influence of the effective interface potential on the shape of the droplets in rectangular and circular capillaries has been usually investigated in two regions \cite{Wong1992,Wong1992b,Ajaev2001,Ajaev2001b,Ajaev2006,Honschoten2010,Starov2010,Mattia2012}. One region corresponds to the droplet surface close to the walls of the channel where the disjoining pressure dominates. The second region corresponds to droplet surface located in the center of the capillary where the effect of disjoining pressure on the shape of the droplet can be ignored. 

In our mesoscopic analysis we investigate the channel heights in the range 10-100 nm and determine the influence of the effective interface potential on droplet shape for any position of the droplet surface. We discuss in detail the geometry of the droplets, such as the thickness of the layer between the droplet and the walls of the channel, as well as the change of the apparent contact angle as function of the increasing height of the channel.

Many of the previous papers on the shapes of the droplets in microchannels have focused on the droplets which were spread between the walls of the channel \cite{Fortes1982,DeSuza2008,Kusumaatmaj2010,Broesch2012}. In the present analysis we put stress on the morphological transition between the sessile state (the droplet touching only one wall of the channel) and the lenticular state (the droplet touching both walls of the channel). The phase diagrams displaying this transition are presented and discussed, in Section II for the macroscopic approach and in Section III for the mesoscopic approach.  In Section IV we point at the role of the line tension when comparing the macroscopic and mesoscopic approaches. We also discuss the solvation force \cite{Butt2009,Dutka2007} which emerges between the channel walls, Section V. It turns out that both in macroscopic  and mesoscopic  approaches the sign of this force changes upon increasing the channel width, turning from repulsive to attractive. We show that the solvation force is zero in situation when the droplet can be inscribed in the circle whose center coincides with the symmetry point of the droplet. Section VI contains discussion.


\section{Macroscopic description}
\begin{figure}[htb]
  \begin{center}
	 \includegraphics[width = \cw \columnwidth]{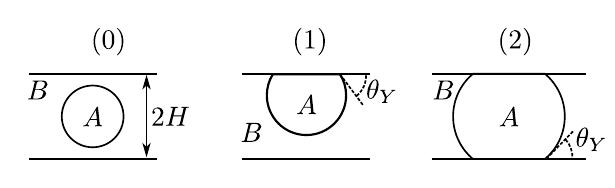}
	 \caption{Three  generic macroscopic states of an $A$-fluid droplet in a planar channel filled with the $B$-fluid. The droplet can touch zero ($0$), one ($1$) or two ($2$) walls, and the macroscopic states are denoted accordingly. The distance between the walls is $2H$. \label{fig_threestates}}
	\end{center}
 \end{figure}

On the macroscopic level one can distinguish three generic equilibrium shapes of the $A$-fluid droplet placed in a flat channel filled with $B$-fluid, see Fig.\,\ref{fig_threestates}. 
For simplicity, we consider a  quasi-two dimensional system which is translationally invariant in one direction. By the shape of the droplet we mean the shape of its cross section perpendicular to the direction in which the system is translationally invariant. Three different morphological states of the droplet can be characterized by the number of walls it remains in contact with: (0), the droplet doesn't touch any of the walls; (1), the droplet touches only one wall, and (2), the droplet touches both walls. In all three cases the shape of the droplet can be described by an arc of a circle. In case $(0)$   the droplet forms a full circle and this state is called the circular state. In the case $(1)$  the shape is a circular segment and we call it the sessile state. The state $(2)$  is called the  lenticular (lens-shaped) state. Whenever the $AB$ interface touches the wall it forms with it an apparent contact angle $\theta_Y$. We note that generically by the contact angle one means the angle between the droplet interface and the wall, i.e. $\pi-\theta_Y$. Here we use the angle $\theta_Y$  to stress that one of the typical experimental realizations of such a system is the channel filled with liquid 
(the $B$-fluid) and its vapor represents the $A$-fluid. So, the angle $\theta_Y$ is the angle formed by the droplet of liquid ($B$) deposited on a planar wall in ambient conditions ($A$). We shall often refer to the  Young's equation 
\beq \label{Young}
 \cos \theta_Y  = \frac{\gamma_{WA}-\gamma_{WB}}{\gamma} \, ,
\eeq
where $\gamma_{WA}$, $\gamma_{WB}$, and $\gamma$ are the wall-$A$ fluid, wall-$B$ fluid, and $A$-fluid - $B$-fluid surface tension coefficients, respectively. We consider the angles $0 \leqslant \theta_Y \leqslant \pi/2$ which is the most common situation in droplet microfluidics \cite{Squires2005,Seemann2012}.

We assume that the fluids $A$ and $B$ are immiscible and incompressible such that the bulk free energy of the system with a droplet  relative to the energy of the channel completely filled by the 
phase $B$  depends neither on the shape nor on the position of the droplet. It depends only on the  cross-sectional area $A$ to which we shall often refer to as the droplet's volume. To track the morphological phase transitions we analyze only the surface free energies, which for the above three states are given by:
\begin{align} \begin{split} \label{omegas}
\Omega_0 &= 2 \gamma \, \sqrt{\pi}\,\sqrt{A} \, ,\\
\Omega_1 &= 2 \gamma \, \sqrt{\pi-\theta_Y+\sin \theta_Y \cos \theta_Y} \, \sqrt{A} \, ,\\
\Omega_2 &= 2 \gamma H \, \Big(\frac{\pi-2\theta_Y}{2 \cos \theta_Y} + \sin \theta_Y +  \frac{1}{2}\cos \theta_Y \frac{A}{H^2} \Big) \, .
\end{split} \end{align}
We notice that the energy of the circular state is always larger than the one corresponding to the sessile state, $\Omega_0>\Omega_1$. There are thus two competing equilibrium states: sessile and lenticular. If however, the circular states are imposed on the system, e.g., via the constraint on the droplet to be placed symmetrically with respect to the center plane of the channel, then one also allows for the circular -- lenticular transition. 

\begin{figure}[htb]
  \begin{center}
	 \includegraphics[width = \cw \columnwidth]{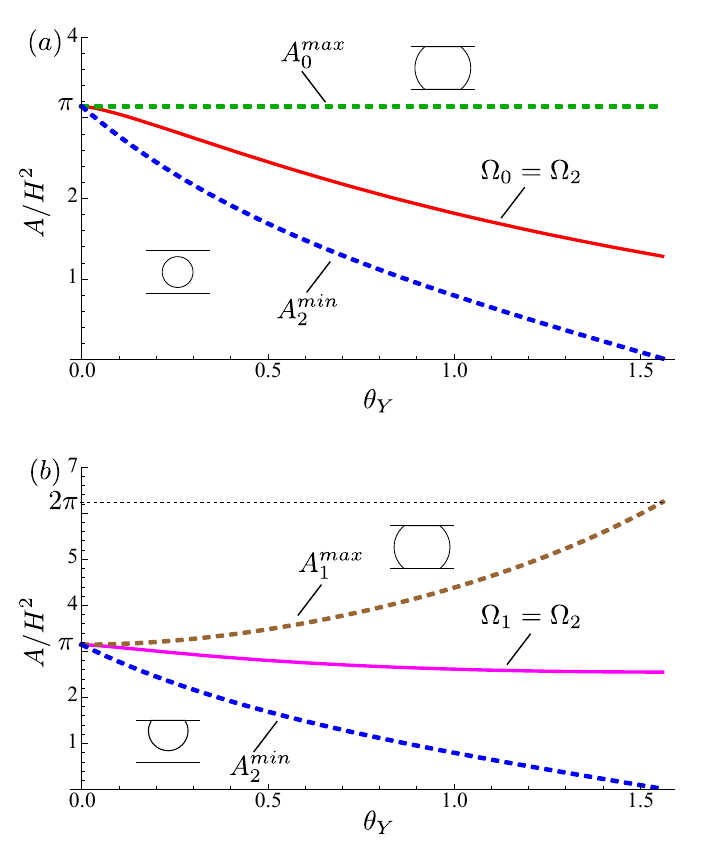}
	 \caption{The phase diagrams illustrating the circular-lenticular $(a)$, and the sessile-lenticular $(b)$ first-order transitions. The phase diagrams are plotted in the contact angle $\theta_Y$ and the volume $A$ variables. The solid lines denote the coexistence curves and the dashed lines are the spinodals. The surface energies ($\Omega_0$, $\Omega_1$, $\Omega_2$) and 
	volumes ($A_0^{max}$, $A_1^{max}$, $A_2^{min}$) are given in Eq.\,(\ref{omegas}) and Eq.\,(\ref{areas}), respectively. \label{fig_macrdiagram}}
	\end{center}
 \end{figure}

\begin{figure}[htb]
  \begin{center}
	 \includegraphics[width = \cw \columnwidth]{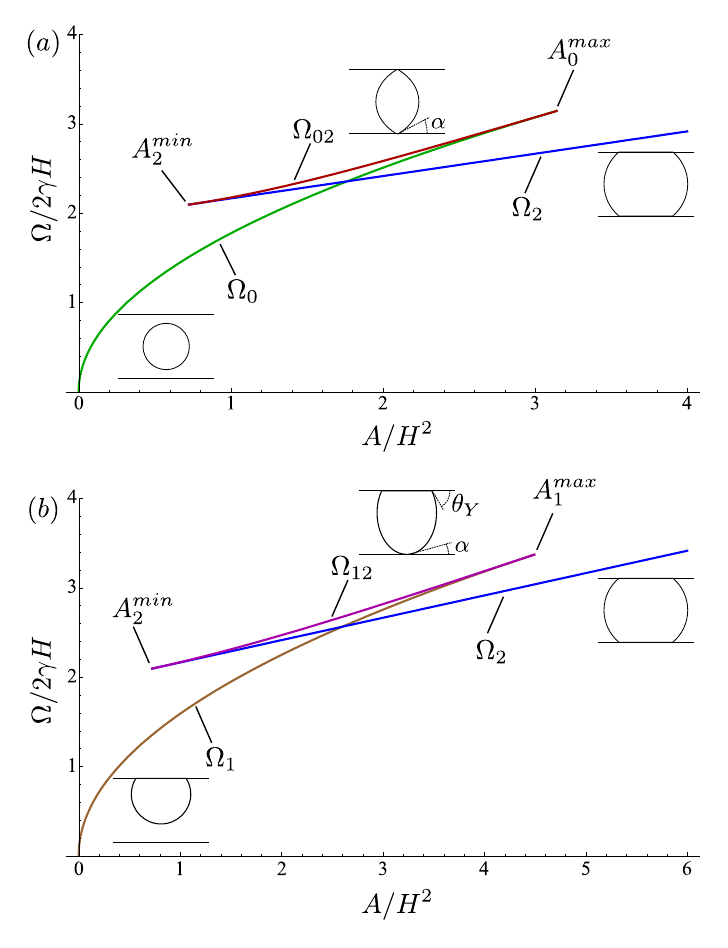}
	 \caption{Surface energy $\Omega$ as a function of an area $A$ for the circular-lenticular $(a)$, and sessile-lenticular $(b)$ transitions for $\theta_Y = \pi/4$. The surface energies ($\Omega_0$, $\Omega_1$, $\Omega_2$) and areas ($A_0^{max}$, $A_1^{max}$, $A_2^{min}$) are given in Eq.\,(\ref{omegas}) and Eq.\,(\ref{areas}), respectively. The symbols $\Omega_{02}$ and $\Omega_{12}$ (Eq.\,(\ref{energyns})) denote the energies of a continuum of unstable states which connect smoothly spinodal points. The angle $\alpha$ characterizing unstable states changes between $0 \leqslant \alpha \leqslant \theta_Y$. \label{fig_macrenergies}}
	\end{center}
 \end{figure}

For large enough volumes $A$ the lenticular state has lowest surface energy, see Figs\,\ref{fig_macrdiagram}, \ref{fig_macrenergies}.
The equations \mbox{$\Omega_0=\Omega_2$} and $\Omega_1=\Omega_2$ determine the coexistence curves. The circular, sessile, and lenticular states cease to exist for areas $A_0>A_0^{max}$, $A_1>A_1^{max}$, $A_2<A_2^{min}$, respectively, which are given by:
\begin{align} \label{areas} \begin{split}
 A_0^{max} =& \pi H^2 \, , \\
 A_1^{max} =& H^2 \frac{4}{(1+\cos \theta_Y)^2}(\pi-\theta_Y+\sin \theta_Y \cos \theta_Y) \, ,\\
 A_2^{min} =& H^2 \Big( \frac{\pi - 2 \theta_Y}{\cos^2 \theta_Y}-2\tan \theta_Y \Big) \, ,
\end{split} \end{align}
and determine the spinodal curves. 

The free energy profiles corresponding to the circular -- lenticular and the sessile -- lenticular transitions are plotted as function of $A/H^2$ for the special choice of $\theta_Y=\pi/4$ in Fig.\,\ref{fig_macrenergies}a and Fig.\,\ref{fig_macrenergies}b, respectively.
On these figures the spinodal points are connected by the lines consisting of particularly constructed unstable states whose morphologies interpolate smoothly between the stable states. The free energies and volumes of these states are denoted by $\Omega_{02}$ and $A_{02}$ for the circular -- lenticular transition, and $\Omega_{12}$ and $A_{12}$ for the sessile -- lenticular transition.  The unstable morphologies can be characterized by only one parameter, the contact angle $\alpha$ which changes from $0 \leqslant \alpha \leqslant \theta_Y$, see Fig.\,\ref{fig_macrenergies}. For $\alpha = \theta_Y$ one has $A_{02}=A_2^{min}$ and $A_{12}=A_2^{min}$ (the lenticular state), while for $\alpha = 0$ the volume $A_{02}=A_0^{max}$ for the circular -- lenticular transition, 
and $A_{12}=A_1^{max}$ for the sessile -- lenticular transition. The free energies of these particular unstable states are given by 
\begin{align} \label{energyns} \begin{split}
\Omega_{02} =& 2 \gamma H \, \frac{\pi-2\alpha}{\cos \alpha} \, ,\\ 
\Omega_{12} =& 2 \gamma H \, \frac{2 (\pi-\theta_Y-\alpha+\cos \theta_Y(\sin \theta_Y -\sin \alpha))}{\cos \theta_Y + \cos \alpha} \, .
\end{split} \end{align}

One could think about other choice of unstable states but these proposed here are characterized by only one parameter and they smoothly interpolate between the spinodal points on Fig.\,\ref{fig_macrenergies}. It turns out that such unstable states appear also in the mesoscopic description.

\section{Mesoscopic description}

\subsection{Equilibrium shape of the droplet}

\begin{center}
\begin{figure}[htb]
  \includegraphics[width = \cw \columnwidth]{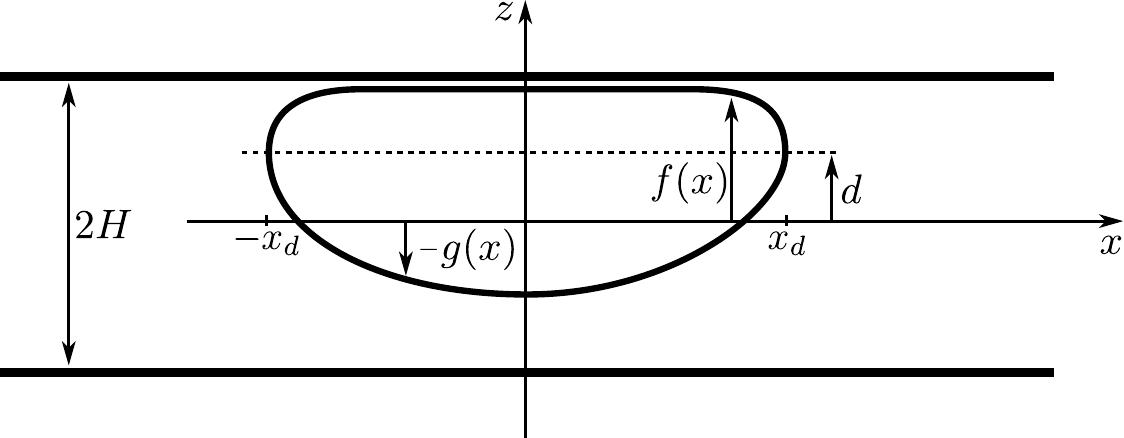}
   \caption{The schematic shape of the $A$-fluid droplet  deposited in a planar channel of height $2H$ and filled with  $B$-fluid. The shape of the droplet is described 
	by two functions: $z=f(x)$ and $z=-g(x)$ which connect smoothly at $f(x_d)=-g(x_d)=d$. We fix the symmetry axis of a droplet at $x=0$. \label{fig_messhape}}
\end{figure}
\end{center}

We assume the system to be translationally invariant in the $y$-direction and due to the invariance of the confining walls in the $x$ direction the equilibrium shape of the droplet has to be symmetric with respect to axis parallel to the $z$-axis. We fix this symmetry axis at $x=0$ and place the walls of the channel at $z=H$, and $z=-H$ (Fig.\,\ref{fig_messhape}).
The shape of a droplet can be described by two functions $z=f(x)$ and $z=-g(x)$ which connect smoothly at $z=d$ with $f'(x_d)=g'(x_d)=-\infty$. The parameter 
$x_d$ is defined implicitly by equation $d=f(x_d)=-g(x_d)$. The surface free energy of the droplet per unit length in $y$-direction equals
\begin{widetext}
\begin{align} \label{ham_drop}
 \begin{split}
\mathscr{H}[f,g] = \int_{-x_d}^{x_d} \dd x \, \Big\{ & \gamma \sqrt{1+(f'(x))^2} +\omega(H-f(x))- \omega(H+f(x)) \\
  & + \gamma \sqrt{1+(g'(x))^2} +\omega(H-g(x))- \omega(H+g(x)) \Big\}\, ,
 \end{split}
\end{align}
\end{widetext}
where $\omega(\ell)$ is the effective interface potential between a flat wall and the interface at a distance $\ell$ from it. The model of this potential stems from the microscopic density functional analysis for the one component fluid in which the attractive parts of the fluid-fluid and wall-fluid interparticle pair potentials are given by \cite{Tasinkevych2006,Tasinkevych2007,Hofmann2010} 
\beq \label{VdWaals}
 w(r) = -\frac{A_F}{(\sigma^2+r^2)^3} \, , \qquad w_W(r)=-\frac{A_W}{(\sigma_W^2+r^2)^3} \, ,
\eeq
where $A_F>0$ and $A_W>0$ are the amplitudes of the interactions while $\sigma$ and $\sigma_W$ are related to the molecular sizes of the fluid and wall particles, e.g., for argon 
$\sigma \approx 0.3\,nm$ \cite{Hansen1969}. For this model the surface tension coefficient is given by \cite{Dutka2012} 
\beq \label{eq_gamma}
\gamma &=& \frac{A_F \pi}{8 \sigma^2}(\rho_B-\rho_A)^2 \, ,
\eeq
and the effective interface potential equals 
\beq
\label{int1}
\omega(\ell) &=& \Delta \rho \frac{\pi}{4} \Big[\frac{\rho_B  A_F}{\sigma^2} \hat \omega(\ell/\sigma) 
   - \frac{\rho_W A_W}{\sigma_W^2} \hat \omega(\ell/\sigma_W) \Big] \, ,
\eeq
where 
\beq \label{omega_p}
	\hat \omega(\ell) &=& 1-\ell \arctan \frac{1}{\ell} \, .
\eeq
Here $\Delta \rho = \rho_B-\rho_A$, and $\rho_A$, $\rho_B$, $\rho_W$ are the $A$-fluid, $B$-fluid, and wall densities, respectively. 
After introducing dimensionless quantities
\beq
\hat \rho =\frac{1}{2} \Big(1-\frac{\rho_A}{\rho_B} \Big) \,, \quad \hat A = \frac{\rho_W A_{W}}{\rho_B A_F} \,, \quad \hat \sigma_W = \frac{\sigma_W}{\sigma}
\eeq
the effective interface potential reduces to \cite{Dutka2012}
\beq
\label{int2}
\omega(\ell) &=& \frac{\gamma}{\hat \rho} \Big[\hat \omega \Big(\frac{\ell}{\sigma}\Big)
  - \frac{\hat A}{{\hat \sigma_W}^2} \hat \omega \Big(\frac{\ell}{\sigma}\frac{1}{\hat \sigma_W}\Big) \Big] \, .
\eeq

The surface tension coefficient and the effective interface potential in  Eqs\,(\ref{eq_gamma}) and (\ref{int1}) can be also obtained from microscopic analysis of the two-component fluid at a planar wall for specific choice of  parameters characterizing  the interparticle interactions, see the Appendix.

The macroscopic Young's contact angle is given by 
\beq
  \cos \theta_Y = 1+\frac{\omega(\ell_\pi)}{\gamma} \, ,
\eeq 
where $\omega(\ell_\pi)$ is the only minimum of the effective interface potential and $\ell_\pi$ is the thickness of the adsorbed layer on a planar substrate. The effective interface potential 
$\omega(\ell \to 0) \to \infty$, and $\omega(\ell \to \infty) \to 0$, so $\omega'(\ell<\ell_\pi)<0$, and  $\omega'(\ell>\ell_\pi)>0$.

As the result of the  minimization of the Hamiltonian under the constraint of the fixed volume $A$ of the droplet
\beq \label{area}
A &=& \int_{-x_d}^{x_d} \dd x \, \Big(\bar f(x)+\bar g(x) \Big)  \, 
\eeq
one obtains the following equations for the equilibrium shape of the droplet $z=\bar f(x)$ and $z=-\bar g(x)$ 
\begin{align} \begin{split} \label{diff_shape}
   \frac{\bar f''(x)}{(1+\bar f'(x)^2)^{3/2}} &=-\bar \omega'(H-\bar f(x))-\bar \omega'(H+\bar f(x))-\lambda \, , \\
	\frac{\bar g''(x)}{(1+\bar g'(x)^2)^{3/2}} &= -\bar \omega'(H-\bar g(x))-\bar \omega'(H+\bar g(x))- \lambda \, ,
\end{split} \end{align}
where $\lambda$ is the Lagrange multiplier, and $\bar \omega(\ell)=\omega(\ell)/\gamma$. After one integration we obtain (we skip the bars)
\begin{align} \begin{split} \label{derivative1}
 \frac{1}{\sqrt{1+f'(x)^2}} &= -\omega(H-f(x))+\omega(H+f(x))+\lambda f(x) +C_1 \, , \\
 \frac{1}{\sqrt{1+g'(x)^2}} &= -\omega(H-g(x))+\omega(H+g(x))+\lambda g(x) +C_2 \, .
\end{split} \end{align}
The boundary conditions:
\begin{align} \begin{split}
  f(x_d) &=-g(x_d)=d \, ,\\
	f'(x=0) &= g'(x=0) = 0, \\
  f'(x_d) &= g'(x_d) = -\infty 
\end{split} \end{align}
 give 
\begin{align} \begin{split} \label{bc1}
 \lambda &= \frac{f_0 \, \eta(f_0)+g_0 \, \eta(g_0)}{f_0+g_0} \, , \\
  C_1 &= -C_2 = \frac{f_0 g_0}{f_0+g_0} \Big(\eta(f_0)-\eta(g_0) \Big) \, , \\
	d &  \eta(d)-1 = \lambda \, d +C_1 \, , 
\end{split} \end{align}
where $f_0 = f(x=0)$, $g_0 = g(x=0)$, and $\eta(z)=(1+\omega(H-z)-\omega(H+z))/z$. In the case of symmetric droplet, i.e. $g_0=f_0$, the parameter $\eta(f_0)$ becomes equal to the Lagrange multiplier. 

Inserting the parameters $\lambda$, $C_1$, $C_2$, and $d$ as function of $f_0$ and $g_0$ into Eq.\,(\ref{derivative1})  
\begin{align} \begin{split}  \label{derivative2}
 \frac{1}{\sqrt{1+f'(x)^2}} &= 1-f(x) \eta(f(x))+\lambda f(x) +C_1 \, , \\
 \frac{1}{\sqrt{1+g'(x)^2}} &= 1-g(x) \eta(g(x))+\lambda g(x) +C_2 \, .
\end{split} \end{align}
one gets the equation 
\beq \label{xd}
 \int_d^{f_0}  \frac{\dd f}{|f'(x)|}  = \int_{-d}^{g_0}  \frac{\dd g}{|g'(x)|} = x_d \, ,
\eeq
which renders the  values of $g_0=g_0(f_0)$. Thus the equilibrium shape of the asymmetric droplet for a fixed volume $A$, Eq.\,(\ref{area}), can be parametrized by one parameter, e.g. $f_0$ which is a function of the volume $f_0=f_0(A)$. 


\subsection{Symmetric droplet}
Typical droplet shapes encountered in droplet microfluidics correspond to the lengths of the droplets which are much larger then the channel height. Then the droplet is symmetric with  respect to the  plane  parallel to the walls and located at the channel's center \cite{Squires2005,Seemann2012}. However, there are situations, e.g. in the flow-focusing method of droplet formation in which the droplet is symmetrically deposited in channel and doesn't touch the sidewalls  \cite{Squires2005}. In this section we discuss the shapes of such symmetric droplets. In particular, we investigate the dependence of the contact angle and the thickness of the films spanned between the walls and the droplet on the channel height.

\begin{center}
\begin{figure}[htb]
  \includegraphics[width = \cw \columnwidth]{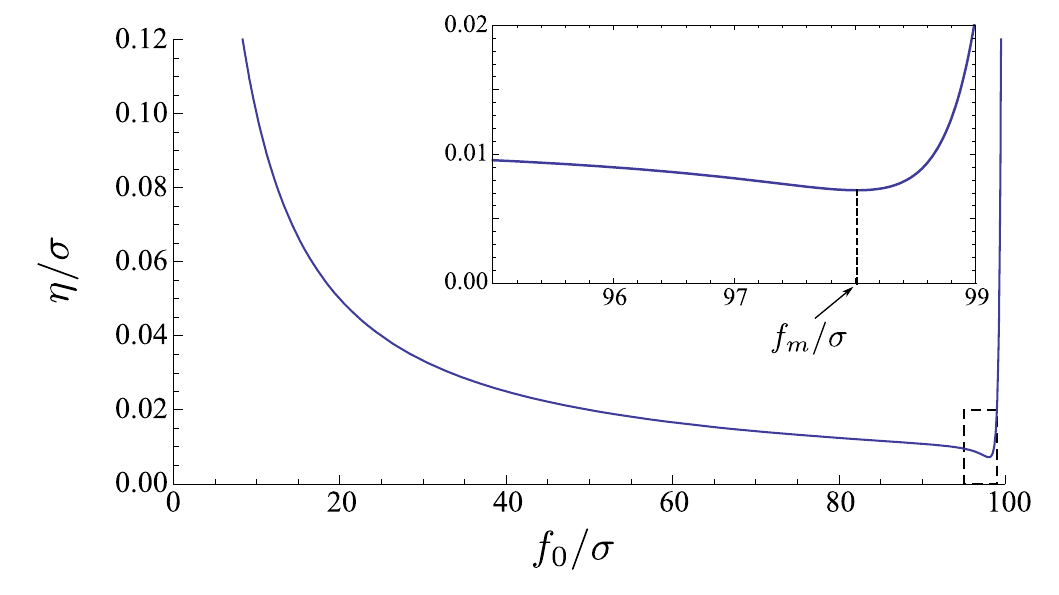}
   \caption{The Lagrange multiplier $\eta(f_0)$ as a function of $f_0$ for $H=100 \sigma$. The inset shows the close-up of $\eta(f_0)$ near its minimum at $f_0=f_m$. The surface tension coefficient and the effective interface potential parameters are chosen such that $\theta=\pi/4$ and $\ell_\pi=2 \sigma$. \label{fig_lambda}}
\end{figure}
\end{center}

\begin{center}
\begin{figure}[htb]
  \includegraphics[width = \cw \columnwidth]{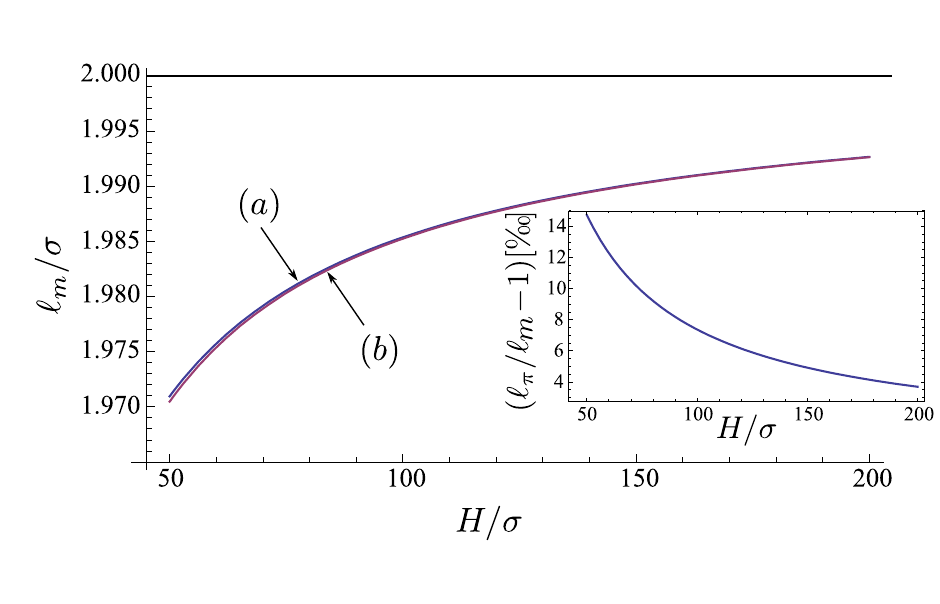}
   \caption{The minimal film thickness $\ell_m$ (curve $(a)$) and its approximation (curve $(b)$), Eq.\,(\ref{ellapp}), as function of $H$ for $A \to \infty$.  The inset shows the relative difference between the $\ell_\pi$ and $\ell_m$. The surface tension coefficient and the effective interface potential parameters are such that $\theta_Y=\pi/4$ and $\ell_\pi=2 \sigma$. \label{fig_ellm}}
\end{figure}
\end{center}

For given macroscopic contact angle $\theta_Y$, Eq.\,(\ref{Young}), and the channel height $2H$, the volume $A$ determines the shape of the droplet. In the symmetric case one has $g_0=f_0$ and Eqs (\ref{area}) and (\ref{derivative2}) give the following relation
\beq \label{eq_Au}
A &=& f_0^{3/2} \int_0^1 \dd t \frac{u(f_0,t)}{\sqrt{(\eta(t f_0)-\eta(f_0)}} \, ,
\eeq
where 
\beq
 u(f_0,t) &=& \sqrt{t}\frac{1-f_0 t \Big(\eta(t f_0)-\eta(f_0)\Big)}{\sqrt{2-f_0 t \Big(\eta(t f_0)-\eta(f_0)\Big)}} \, .
\eeq
The function $u(f_0,t)$ is finite for $t \in [0,1]$. The Lagrange multiplier $\eta(f_0)$ is positive for $f_0>0$ and has a minimum at $f_0 = f_m$ denoted as $\eta_m=\eta(f_0=f_m)$, 
see Fig.\,\ref{fig_lambda}.
For $t \to 1$ 
\begin{align} \begin{split}
  \eta(t f_0)-\eta(f_0) = & -f_0 \eta'(f_0)(1-t) \\ 
	  & +\frac{1}{2} f_0^2 \eta''(f_0)(1-t)^2 + \ldots \, ,
\end{split} \end{align} 
and it follows from Eq.\,(\ref{eq_Au}) that $A \to \infty$ for $f_0 \to f_m$. Thus for given height of the channel, the minimal thickness of the film between the droplet and the wall $\ell_m = H -f_m$ is attained as $A \to \infty$. One can check that $\ell_m<\ell_\pi$, where $\ell_\pi$ fulfills $\omega'(\ell_\pi)=0$, and for increasing  $H$ the minimal film thickness behaves as 
\beq \label{ellapp}
\ell_m &=& \ell_\pi - \frac{\cos \theta_Y }{\omega_\pi''(\ell_\pi)} \,  \frac{1}{H} + \mathcal{O}\Big(\frac{1}{H^2}\Big)\, , 
\eeq
see Fig.\ref{fig_ellm}. 
   
\begin{center}
\begin{figure}[htb]
	\includegraphics[width = \cw \columnwidth]{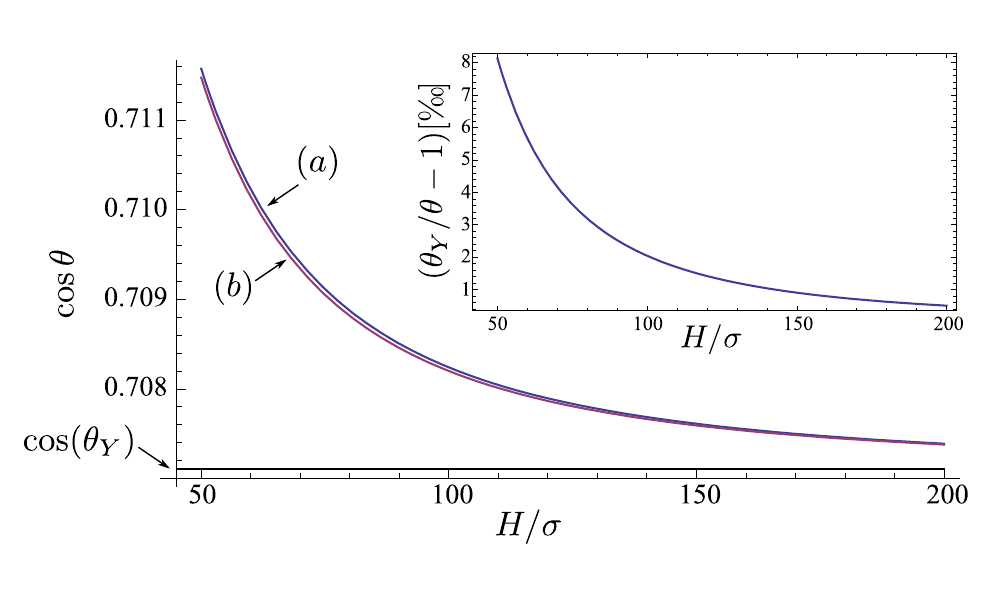}
   \caption{The cosine of mesoscopic contact angle $\theta$ (curve $(a)$) and its approximation (curve $(b)$), Eq.\,(\ref{cosapp}), as function of $H$ for $A \to \infty$. The inset shows the relative difference between the contact angle of the droplet and the Young's angle. The surface tension coefficient and the effective interface potential parameters are such that $\theta_Y=\pi/4$ and $\ell_\pi=2 \sigma$. \label{fig_cosapp}}
\end{figure}
\end{center}

Macroscopically, the apparent contact angle of the symmetric droplet in the lenticular state is given by 
\beq 
  \cos \theta_Y = \frac{H}{R} \, ,
\eeq
where $R$ is the radius of curvature of the droplet, and $H$ is at the same time one half of the channel' height and the highest position of the droplet interface. In mesoscopic description, we define the contact angle in the same way, and as the radius of curvature we take the inverse of a curvature at $z=0$
\begin{align} \begin{split} \label{mes_angle}
  \cos \theta &= f_0 \Big(\eta(f_0)+2\omega'(H)\Big) \\ & = 1+\omega(H-f_0)-\omega(H+f_0)+2\omega'(H)f_0 \, .
\end{split} \end{align}
The effective interface potential $\omega(z) \propto 1/z^2$ for $z \gg 1 $, so for large $H/\sigma$ and $A \to \infty$ the mesoscopic contact angle behaves as
\begin{align} \begin{split} \label{cosapp}
  \cos \theta =& 1+\omega(\ell_m)-\omega(2H-\ell_m)+2\omega'(H)(H-\ell_m) \\ 
	 =& \cos \theta_Y +\frac{\cos^2 \theta_Y }{2 \omega_\pi''(\ell_\pi)}\frac{1}{H^2} \\ & \quad -\omega(2H)+2\omega'(H)H + \mathcal{O}\Big(\frac{1}{H^3}\Big) \, ,
\end{split} \end{align}
see Fig.\,\ref{fig_cosapp},


\subsection{Morphological transition} 
\begin{center}
\begin{figure}[htb]
  \includegraphics[width = \cw \columnwidth]{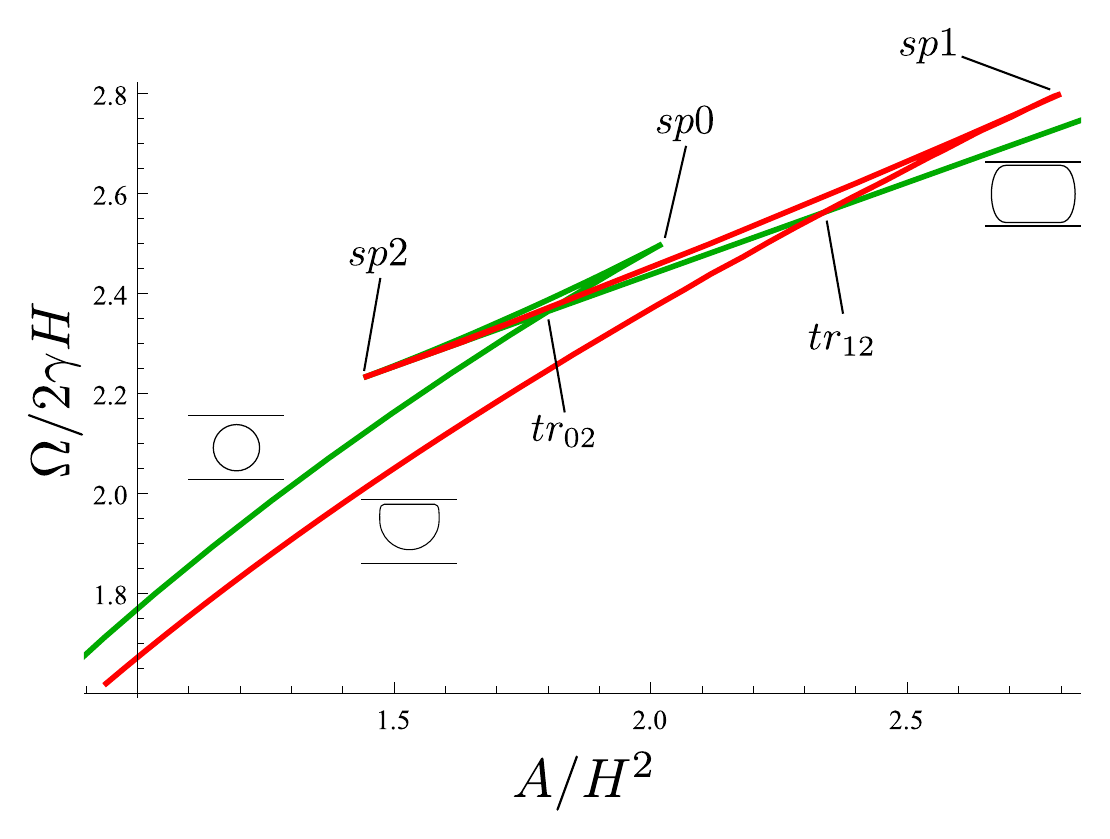}
   \caption{Surface energy $\Omega$ as function of volume $A$ for symmetric (green line) and asymmetric (red line) shapes for $\theta_Y = \pi/4$. The points $tr_{02}$, $tr_{12}$, $sp0$, $sp1$, and $sp2$ denote the transition and spinodal points. The surface tension coefficient and the effective interface potential parameters are such that  $\ell_\pi=2 \sigma$, and $H=50 \sigma$. \label{fig_mesenergy}}
\end{figure}
\end{center}

The free energy profile $\Omega$ as  function of volume $A$ obtained in the mesoscopic analysis is shown on Fig.\,\ref{fig_mesenergy}.
The points marked with: $tr_{02}$, $tr_{12}$, $sp0$, $sp1$, and $sp2$ denote the transition and spinodal points. The circular, sessile and lenticular states cease to exist for points $(A_{sp0},\Omega_{sp0})$, $(A_{sp1},\Omega_{sp1})$, and $(A_{sp2},\Omega_{sp2})$, respectively. The lines connecting the spinodal points correspond to unstable states. We notice that for $A>A_{sp1}$ the droplets of asymmetric shapes cannot exist in a flat channel. 

\begin{center}
\begin{figure}[htb]
  \includegraphics[width = \cw \columnwidth]{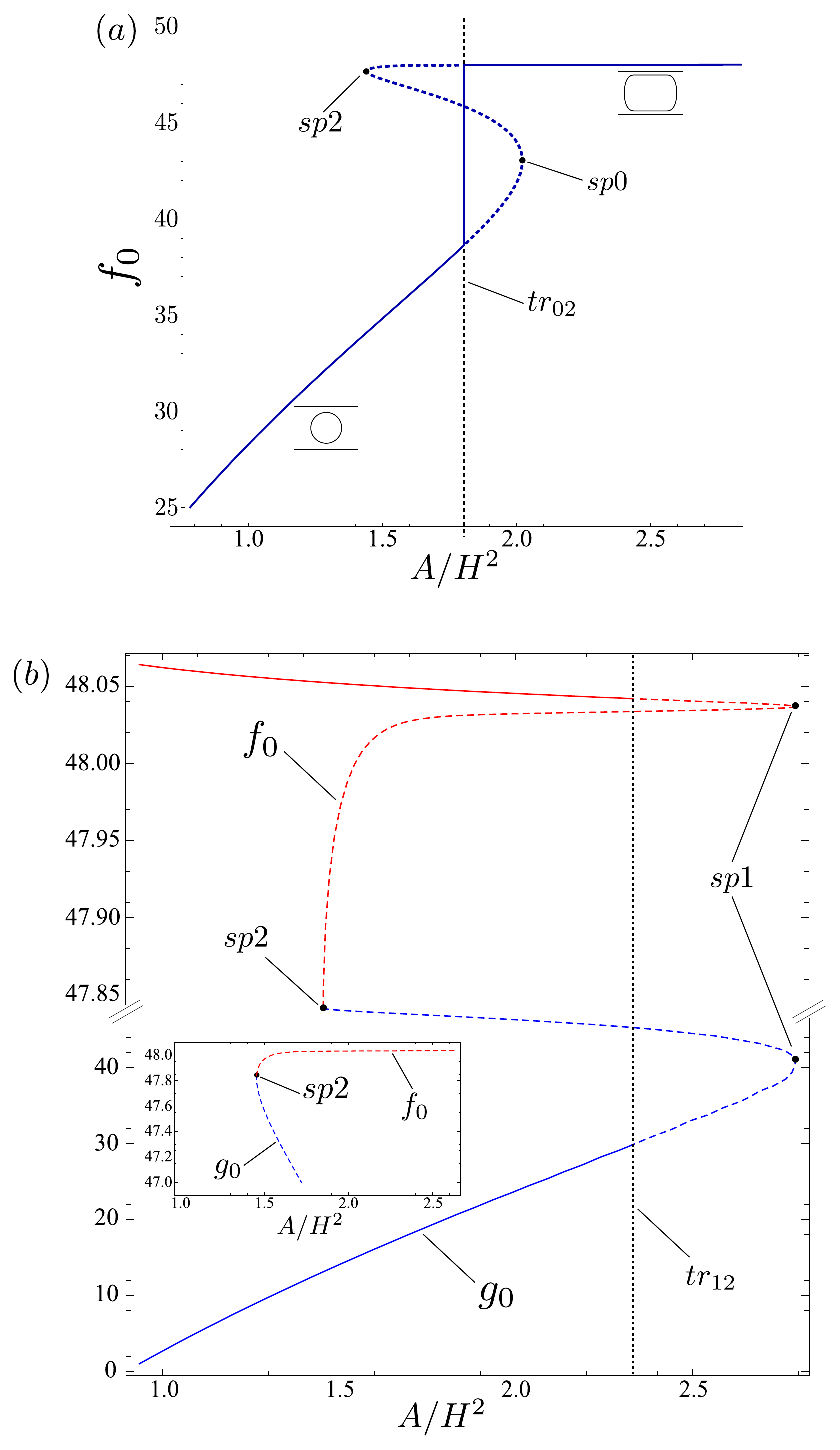}
   \caption{The highest $f_0$ and the lowest $g_0$ positions of the droplet surface as function of the volume $A$ for circular -- lenticular, $(a$), and sessile -- lenticular, $(b)$, transitions. In $(b)$ we do not display the lenticular branch which is the same as in $(a)$. The solid lines correspond to stable states, and dashed lines to the metastable and unstable states. The transition points are marked with $tr_{02}$, $tr_{12}$, and the spinodal points with $sp0$, $sp1$ and $sp2$ at which the dashed lines connect smoothly, see the inset in $(b)$. The surface tension coefficient and the effective interface potential parameters are such that $\theta_Y=\pi/4$ and $\ell_\pi=2 \sigma$, and $H=50\sigma$. \label{fig_mes_f0g0}}
\end{figure}
\end{center}

\begin{center}
\begin{figure}[htb]
  \includegraphics[width = 0.7 \columnwidth]{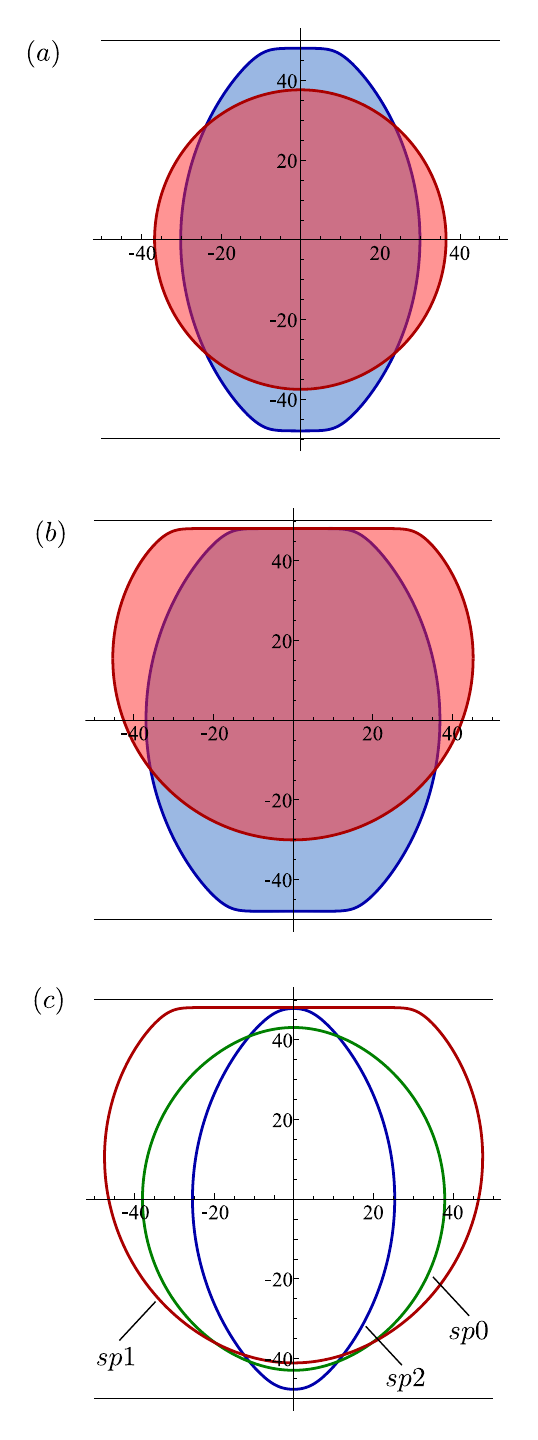}
   \caption{The equilibrium shapes of the droplets in the circular -- lenticular, $(a)$, and the sessile -- lenticular, $(b)$, transitions at $A=A_{tr02}$ and $A=A_{tr12}$, respectively. Part $(c)$ shows the droplet shapes at spinodal points $sp0$, $sp1$, and $sp2$. The surface tension coefficient and the effective interface potential parameters are such that $\theta_Y=\pi/4$ and $\ell_\pi=2 \sigma$, and $H=50\sigma$. 
	\label{fig_mestransshape}}
\end{figure}
\end{center}

The droplet profiles fulfill Eq.\,(\ref{derivative1}) and can be parametrized by $f_0$, and $g_0$ -- the highest and lowest position of the interface; for symmetric droplets $g_0=f_0$. For both the circular -- lenticular, and the sessile -- lenticular transitions the stable and metastable states are characterized by an increasing $f_0$ and $g_0$ as a function of $A$, Fig.\,\ref{fig_mes_f0g0}.
 We notice that  as soon as the lenticular state is attained the parameter $f_0$ remains practically constant; it doesn't increase more than $0.1 \sigma$, see also Fig.\,\ref{fig_ellm}. Upon increasing the volume of the droplets $A$ there is a jump in $f_0$ and $g_0$  at the transition points, Fig.\,\ref{fig_mestransshape}.
In the circular state the distance between the droplet surface and the wall is large enough such that the effective interaction between the wall and the droplet surface has no effect on the shape of the droplet. In mesoscopic analysis the situation in which the droplet surface is within the distance $\ell \approx \ell_\pi$ to the wall corresponds to the droplet-wall contact in the macroscopic description. At spinodal point $sp0$ the minimal film thickness between the droplet in the circular state and the wall is much larger than $\ell_\pi$; also for the sessile state $sp1$ the distance $H -g_0 \gg \ell_\pi$, Fig.\,\ref{fig_mestransshape}. For spinodal point $sp2$ the mesoscopic shape resembles its macroscopic counterpart.

\cleardoublepage

\section{Line tension}

 We have already noticed that the mesoscopic circular states have the same energy as the macroscopic ones. On the other hand the mesoscopic sessile and lenticular states have a lower free energy as compared to their macroscopic counterparts. The mesoscopic free energy, beside the contribution scaling with the surface of the droplet, contains also a line contribution connected with two (in the sessile state) and four (in the lenticular state) three-phase contact lines extending in the $y$-direction \cite{Schimmele2007}. This line contribution was not taken into account in the macroscopic description. For long-ranged van der Waals forces rendering continuous wetting transition and exploited in our analysis the line tension coefficient is negative \cite{Getta1998}.  The formula for the line tension coefficient contains various contributions among which the most significant one includes the interaction of the solid wall with the interface detaching from the wall. According to Eq.\,(4.4) in Ref.\,\onlinecite{Getta1998} it takes the form
\beq \label{eq_linetension}
 \tau = \frac{1}{\tan \theta_Y} \int_{\ell_\pi}^\infty \omega(y) \, ,
\eeq
and contributes roughly to one half of the value of the line tension coefficient. Although the authors in Ref.\,\onlinecite{Getta1998} analyzed the behavior of line tension in the vicinity of wetting temperature, where $\theta_Y \to 0$,  we use $2 \tau$ as the estimate of the line contribution to the free energy stemming from a single three-phase contact line, also away from the wetting point. 

\begin{center}
\begin{figure}[htb]
  \includegraphics[width = \cw \columnwidth]{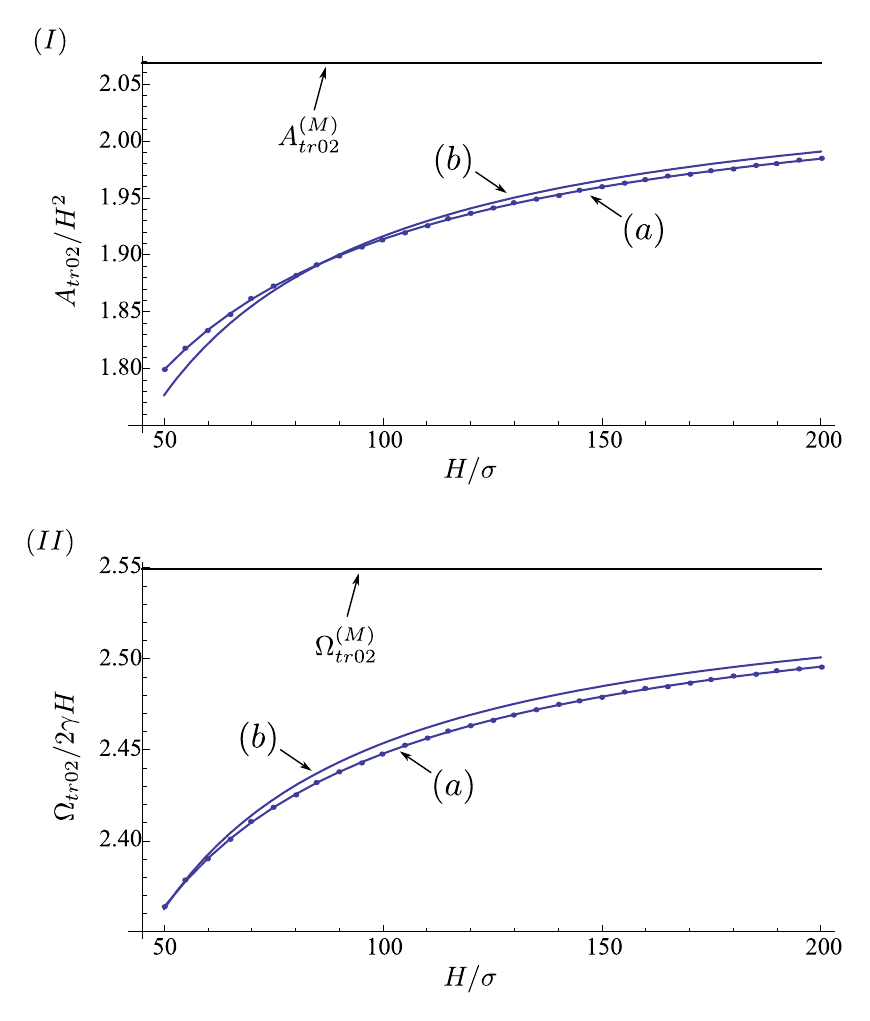}
   \caption{The volume $(I)$ and free energy $(II)$ of the circular -- lenticular transition as function of  $H$ obtained within mesoscopic description, $(a)$, and macroscopic description   including the line tension contributions, $(b)$. The values of the volume and  macroscopic free energy at morphological phase transition  without taking into account the line tension are marked with superscripts $(M)$. \label{fig_transitionH}}
\end{figure}
\end{center}

\begin{center}
\begin{figure}[htb] 
  \includegraphics[width = \cw \columnwidth]{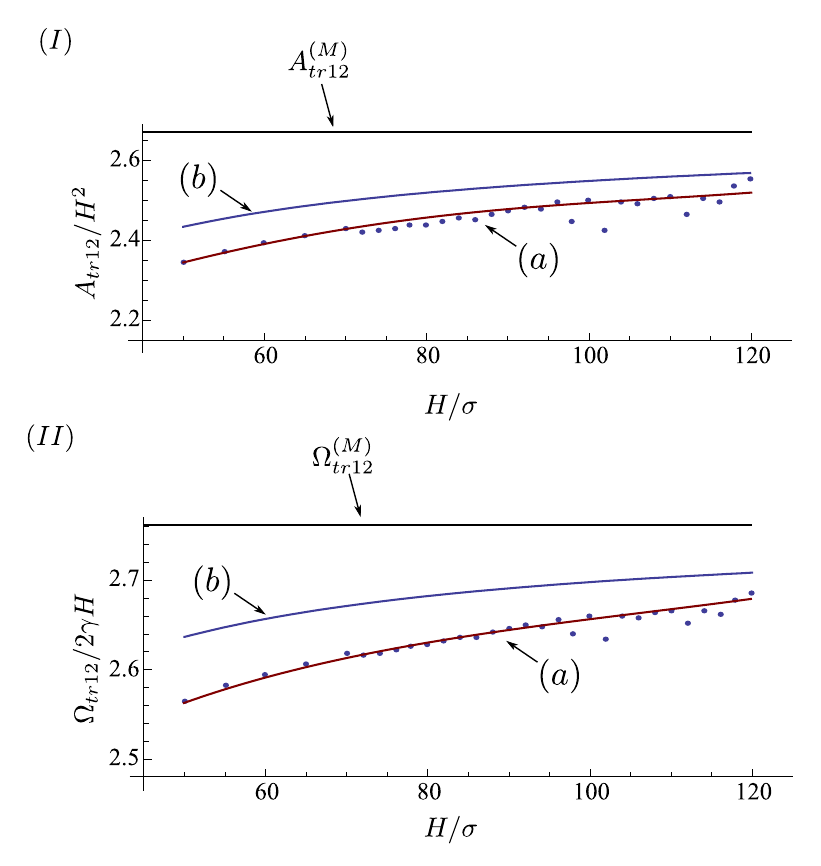}
   \caption{The volume $(I)$ and free energy $(II)$ of the sessile -- lenticular transition as  function of  $H$ obtained within mesoscopic description, $(a)$, and macroscopic description including the line tension contributions, $(b)$. The values of the volume and the macroscopic free energy at morphological phase transition including the line tension are marked with superscripts $(M)$. \label{fig_transitionH_asym}}
\end{figure}
\end{center}

The inclusion of the line tension into the macroscopic description results in the change of the value of the volume $A$ at which the morphological transitions take place. It can be calculated by solving equations 
\begin{align} \begin{split}
  \Omega_0(A) &= \Omega_2(A)+ 8 \tau \\
	\Omega_1(A)+4 \tau & = \Omega_2(A)+ 8 \tau
\end{split} \end{align}
for the circular -- lenticular, and the sessile -- lenticular transitions, respectively. In the case of the circular -- lenticular transition the values of  $A_{tr}$ and  $\Omega_{tr}$ characterizing the morphological phase transition calculated within the mesoscopic description are well approximated by the values obtained within the macroscopic description with the inclusion of the line tension contributions, Fig.\,\ref{fig_transitionH}. 
In the case of the sessile -- lenticular transition this procedure leads to results presented on  Fig.\,\ref{fig_transitionH_asym}. The line tension calculated in the full mesoscopic description turns out to be smaller (more negative) than the approximate value $2 \tau$  used within this simple approach. 

\begin{center}
\begin{figure}[htb]
  \includegraphics[width = \cw \columnwidth]{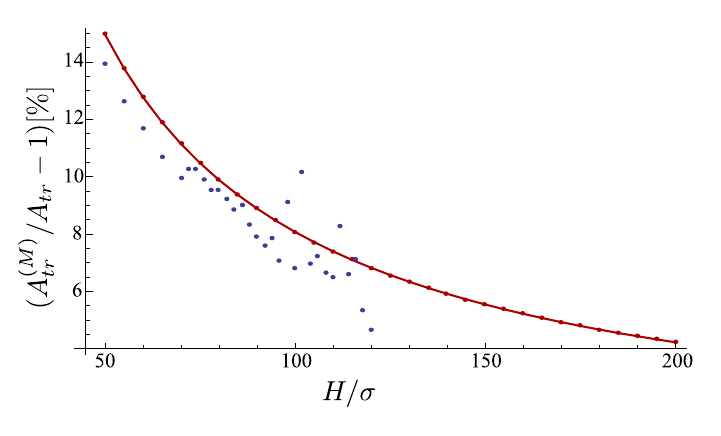}
   \caption{The relative difference between the volume at the circular -- lenticular (red dots), and the sessile -- lenticular (blue dots)  transitions within macroscopic approach without taking into account the line tension, $A_{tr}^{(M)}$, and mesoscopic approaches, $A_{tr}$.   \label{fig_percent}}
\end{figure}
\end{center}

The parameter $f_0$ characterizing the lenticular state, and therefore the contact angle $\theta$ (Eq.\,(\ref{mes_angle})), remain practically independent of the area $A$. For the values of the thermodynamic and geometric parameters considered in our analysis, and for  $H>50\sigma \approx 15 \, nm$, Fig.\,\ref{fig_cosapp}, the Young's contact angle is a very good approximation of the mesoscopic contact angle. The relative difference is smaller than one per million. The line tension coefficient makes  between $3\% - 0.5 \%$ of the total free energy for $H$ within $50 \sigma - 200 \sigma$. One could thus expect that the macroscopic description without including the line tension contributions would predict the values of the volume at the phase transition to be located within similar error margin, i.e. below $3\%$. However, this is not the case and the difference between the macroscopic and mesoscopic description is more pronounced, between $5\% - 14 \%$, Fig.\,\ref{fig_percent}.
The relative difference $A_{tr}^{(M)}/A_{tr}-1$ between the volume at the circular -- lenticular transition within macroscopic description without including the line tension contributions 
($A_{tr}^{(M)}$) and mesoscopic description ($A_{tr}$) decreases like $1/H$, as expected. In case of the sessile -- lenticular transition the numerical results are less reliable due to numerical errors induced by solving Eq.\,\ref{xd} and calculating the droplet shape and its free energy in the sessile state.


\section{Solvation force}
Insertion of the $A$-fluid droplet into the channel filled by the $B$-fluid changes the free energy of the system and, in particular, modifies the solvation force acting between the sidewalls. The solvation force $F$ is calculated as 
\mbox{$F= - \partial \Omega / \partial (2H)$} at fixed volume of the droplet.
  
\begin{center}
\begin{figure}[htb]
  \includegraphics[width = \cw \columnwidth]{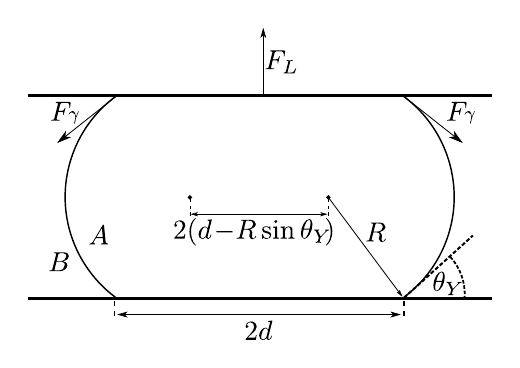}
   \caption{Schematic shape of the droplet in macroscopic description of the lenticular state. The length of the droplet-wall interface is denoted by $2d$, the radius of curvature of the $A-B$ interface by $R$, and the contact angle by $\theta_Y$. The solvation force (per unit length in the $y$-direction) acting on the upper wall contains contribution from the  Laplace force $F_L = 2d \Delta p = 2d \gamma /R$, where $\Delta p$ is the Laplace pressure, and the surface tension contribution denoted by  $F_\gamma = \gamma$. \label{fig_sily_makr_sketch}}
\end{figure}
\end{center}

\begin{center}
\begin{figure}[htb]
  \includegraphics[width = \cw \columnwidth]{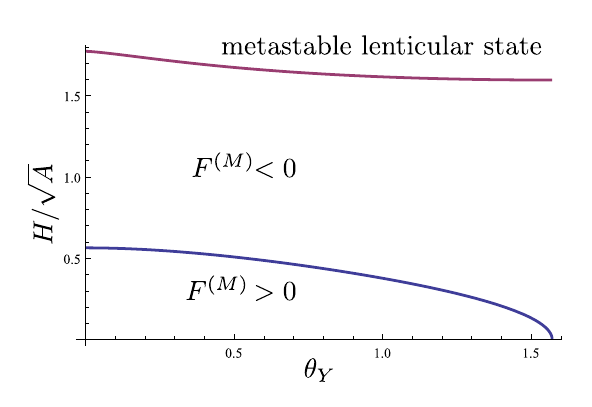}
   \caption{The diagram in ($\theta_Y, H$) variables illustrating the change of sign of the solvation force in the case of macroscopically analyzed lenticular state at fixed volume $A$. The red line denotes the sessile-lenticular transition above which the lenticular state is metastable. \label{fig_silymakro}}
\end{figure}
\end{center}

In macroscopic description only the free energy of the lenticular state depends on the channel height, Eq.\,(\ref{omegas}). The solvation force per unit length in the $y$-direction is thus non-zero and equals \cite{Kusumaatmaj2010}
\begin{align}
 \begin{split}
 F^{(M)} =& - \frac{\partial \Omega_2}{\partial (2H)} = \gamma \, \Big(\frac{1}{2}\cos \theta_Y \frac{A}{H^2}-\frac{\pi-2\theta_Y}{2 \cos \theta_Y} - \sin \theta_Y \Big) \\
    = & \frac{\gamma}{R} \Big(2d - 2R \sin \theta_Y \Big) = 2 d \Delta p - 2 \gamma \sin \theta_Y \ , 
\end{split}
\end{align}
where $2d$ is the length of the droplet-wall interface, $R$ is the radius of curvature of $AB$ interface, and \mbox{$\Delta p = p_A-p_B=\gamma /R$} is the Laplace pressure, 
see Fig.\,\ref{fig_sily_makr_sketch}. 
Accordingly, the solvation force is positive when the center of each arc of the circle forming the $AB$ interface (points $O_1$ and $O_2$ in Fig.\,\ref{fig_sily_makr_sketch}) and the corresponding interface  are located on the same side of the droplet symmetry axis perpendicular to the channel walls. Upon increasing  the channel height the solvation force decreases and becomes zero when 
the arcs centers $O_1$ and $O_2$ merge on the symmetry axis, and becomes negative for larger values of $H$, Fig.\,\ref{fig_silymakro}. We notice that - for larger values of $H$ - the lenticular state becomes metastable against the sessile  or circular state. In particular, for large enough  channel height one has $2d = 0$, and the lenticular state ceases to exist. 

\begin{center}
\begin{figure}[htb]
  \includegraphics[width = \columnwidth]{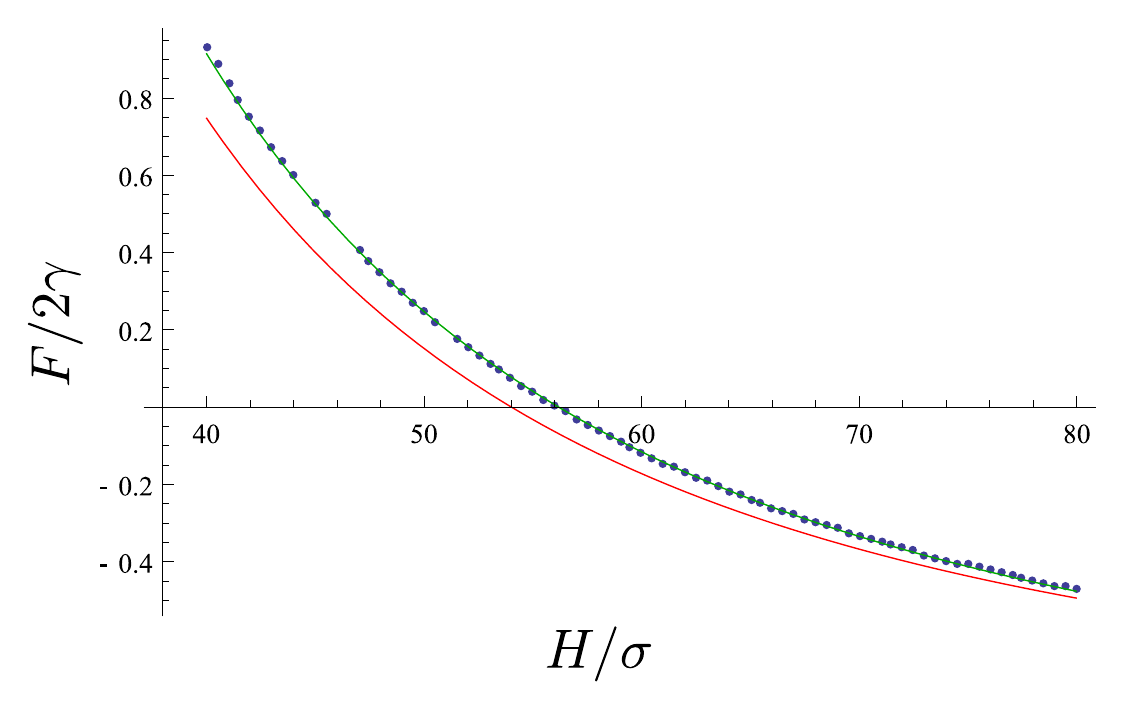}
   \caption{The solvation force $F$ as a function of $H$ calculated within mesoscopic  (dots) and macroscopic (red line) analysis. The green line is introduced to guide the eye. The calculation was done for $A=15000\sigma^2$; the surface tension coefficient $\gamma$ and the effective interface potential parameters are such that $\theta_Y=\pi/4$ and $\ell_\pi=2 \sigma$. \label{fig_sily_mezo}}
\end{figure}
\end{center}

\begin{widetext}
In mesoscopic description, the free energy of the lenticular state, Eq.\,(\ref{ham_drop}), is given by 
\beq
  \mathscr{H}[\bar f] = 4 \int_{0}^{x_d} \dd x \, \Big\{ \gamma \sqrt{1+(\bar f'(x))^2} +\omega(H-\bar f(x))- \omega(H+\bar f(x))\Big\} \, , 
\eeq
where $z=\bar f(x)$ describes the equilibrium shape of the droplet. Correspondingly, the solvation force is given by 
\begin{align} \label{eq_Fmeso}
 \begin{split}
 F = & - \frac{\partial \mathscr{H}[\bar f]}{\partial (2H)} = - \frac{1}{2}\frac{\partial \mathscr{H}[\bar f]}{\partial H} \\
  = &  - 2 \int_{0}^{x_d} \dd x \, \Big\{ \omega'(H-\bar f(x))- \omega'(H+\bar f(x)) \Big\} \, . 
\end{split}
\end{align}
\end{widetext}
In agreement with the macroscopic analysis conclusions the solvation force is positive for small values of the channel height and becomes negative for larger values, Fig.\,\ref{fig_sily_mezo}. For decreasing  $H$ the thickness of the film between the droplet and the wall, $\ell_0=H-f_0$ decreases and the derivative $\omega'(\ell_0)$ becomes more negative; therefore the solvation force can be positive. For higher values of $H$ the thickness $\ell_0$ increases, $\omega'(\ell_0)$ becomes less negative and the solvation force changes its sign.  


\begin{center}
\begin{figure}[htb]
  \includegraphics[width = 0.8\columnwidth]{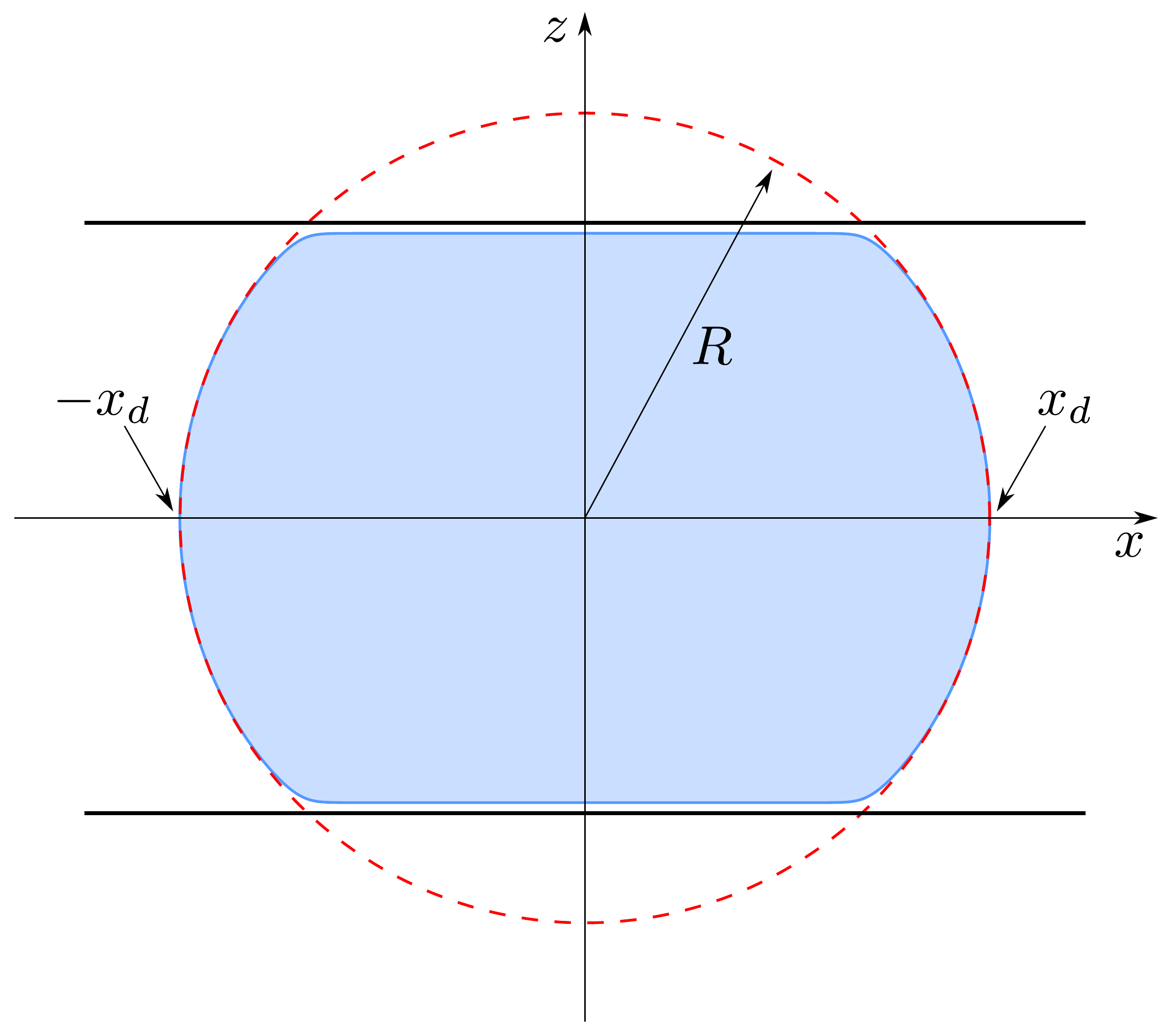}
   \caption{The shape of the mesoscopic droplet corresponding to zero solvation force, $F=0$. The shape of the droplet is such that the radius of curvature of the interface $R$  at $z=0$ equals $x_d$, and the $A-B$ interface can be approximated by two arcs of the same circle with the center at $(x=0,z=0)$. The droplet shape corresponding to $F^{(M)}=0$ in the macroscopic analysis also exhibits this feature. \label{fig_wFkolo2}}
\end{figure} 
\end{center}

The shape of the droplet corresponding to $F=0$ is such that the radius of curvature $R$ of the droplet at $z=0$ equals $x_d$, $R=x_d$. In this situation, the $A-B$ interface can be approximated by two arcs of the same circle with the center at $(x=0,\,z=0)$, Fig.\,\ref{fig_wFkolo2}.

\clearpage
\section{Discussion}

We have derived the phase diagrams for the circular-lenticular and the sessile-lenticular morphological transitions of a droplet in a channel within two approaches: macroscopic and mesoscopic. Since the free energy of the sessile state is always smaller than that of the circular state the former transition can be observed only when droplet configurations which are symmetric with respect to the center plane of a channel are imposed on the system. e.g., via the appropriate constraint. Both morphological transitions are first-order and are accompanied by the presence of metastable and unstable states. In mesoscopic description the free energy profile, Fig.\,\ref{fig_mesenergy}, is qualitatively the same as in the macroscopic description, Fig.\,\ref{fig_macrenergies}. However, the macroscopic approach which is not corrected by the inclusion of the line tension contributions, overestimates both the free energies and volumes at transition points up to 14\% as compared to the mesoscopic values, see Fig.\,\ref{fig_percent}. This comparison can be substantially improved by including the contact angle dependent line tension coefficient, Figs.\,\ref{fig_transitionH}, \ref{fig_transitionH_asym}.    

The long-ranged interparticle interactions taken into account in analysis, Eq.\,(\ref{VdWaals}), render the critical wetting transition at a planar substrate and lead to negative line tension coefficient. The interparticle interactions leading to the first-order wetting transition give positive line tension coefficient \cite{Indekeu1992,Schimmele2007}. We suppose that in this case the values of volumes  characterizing the sessile --lenticular transition will be larger than in the case of negative line tension coefficients. In addition to the  the transition points, also the spinodal points would change within the macroscopic description including the line tension contributions. Thus the analysis of the droplet states in the nanochannels can give us a hint about the underlying interparticle interaction and the order of the wetting transition.

In the mesoscopic description there is always a layer of the host $B$-fluid separating the $A$-fluid droplet from the channel walls. This is the most profound difference between the mesoscopic description and its macroscopic counterpart, where one allows for the droplet-wall interface. 
Nevertheless, also in the mesoscopic approach one can define the contact angle $\theta$, see Eq.\,(\ref{mes_angle}). This angle approaches the macroscopic Young's angle $\theta_Y$ for $H \to \infty$ and droplet's volume $A \to \infty$. For mesoscopic channel heights and  large droplets ($A \to \infty$) this angle is smaller than $\theta_Y$, Fig.\,\ref{fig_cosapp}. The difference $\theta_Y-\theta$ decreases with increasing height and its relative value is smaller than one per mil already for $H=50\sigma$.

In the mesoscopic description of the lenticular states of large droplets the film thickness between the droplet and the wall $\ell_0=H-f_0$ is smaller than $\ell_\pi$, i.e., the thickness of the adsorption layer of the $B$-fluid on a planar substrate, Fig.\,\ref{fig_ellm}. The difference $\ell_\pi-\ell_0$ decreases with increasing channel height and for $H>50\sigma$ it is smaller than $0.05 \sigma$. However, even this minor difference give rise to the positive (repulsive) solvation force, which is also present in macroscopic description. Approximating $\ell_0$ by $\ell_\pi$ would  incorrectly render the always  negative (attractive) solvation force, see Eq.\,(\ref{eq_Fmeso}). 

The predicted change of sign of the solvation force in the lenticular state, also reported in Refs.\,\onlinecite{Kusumaatmaj2010,DeSuza2008}, brings new issue in experimental micro- and nanofluidics. Suppose that one wall of the channel filled with the $B$-fluid can move in the direction perpendicular to it. Inserting many identical droplets of the $A$-fluid of fixed volume (with large enough distance between them to prevent their coalescence)  will determine the distance between the walls of the channel. This height is a function of number and the volume of the inserted droplets. Generally, the droplets of the $A$-fluid  immersed in the channel filled with the $B$-fluid can act as micro- or nanodampeners (shock absorbers).  

\appendix*
\section{Effective interaction between a flat wall and  droplet surface \label{app_interaction}}
\begin{figure}[htb]
  \begin{center}
	 \includegraphics[width = \cw \columnwidth]{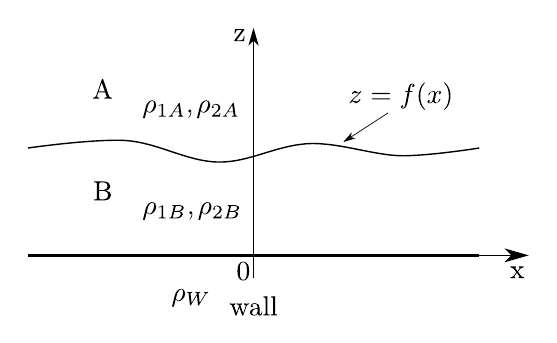}
	 \caption{The two-component system at a planar substrate. The system is invariant in $y$-direction and $z=f(x)$ is a fluid-fluid interface separating phases $A$, and $B$ rich in component $1$ and $2$, respectively. \label{fig_mixture}}
	\end{center}
 \end{figure}

Consider an interface fluctuating near a planar wall, see  Fig.\,\ref{fig_mixture}. This interface separates the phases $A$ and $B$ rich in components $1$ and $2$, respectively. The thermodynamic state of the system corresponds to the coexistence of these $A$ and $B$ phases of the binary mixture.  The system is invariant in $y$-direction and $z=f(x)$ denotes the position of the interface. 
The interfacial Hamiltonian takes the form \cite{Dietrich1991,Hiester2006} 
\begin{align}
\mathscr{H}_{AB}[f] = \int_{-\infty}^\infty \dd x \Big\{ \gamma_{AB} \sqrt{1+(f'(x))^2} +\omega_{AB}(f(x)) \Big) \, 
\end{align}
where $\gamma_{AB}$ is the surface tension coefficient, and  $\omega_{AB}(\ell)$ is the effective interface potential between the wall and the interface located at the distance $\ell$ from it. We consider the following model of long-ranged attractive interparticle $w_{ij}(r)$ and wall-particle $w_{iW}(r)$ interactions  
\begin{align}
 w_{ij}(r) = -\frac{A_{ij}}{(\sigma_{ij}^2+r^2)^3} \, , \qquad w_{iW}(r)=-\frac{A_{iW}}{(\sigma_{iW}^2+r^2)^3} \, ,
\end{align}
where $i,j=1,2$ enumerate the fluid components. The amplitudes $A_{ij}$and $A_{iW}$ are positive; the positive parameters $\sigma_{ij}$ and $\sigma_{iW}$ are related to the molecular sizes of the fluid and substrate particles. For this model the surface tension coefficient is equal
\beq 
\gamma_{AB} &=& \frac{\pi}{8} \sum_{i,j=1}^2 \frac{A_{ij}}{\sigma_{ij}^2}(\rho_{iB}-\rho_{iA})(\rho_{jB}-\rho_{jA}) \, ,
\eeq
and the effective interface potential 
\begin{align} \begin{split}
\omega_{AB}(\ell) =& \frac{\pi}{4} \, \sum_{i,j=1}^2 (\rho_{iB}-\rho_{iA})\, \\ & \qquad
 \Big(\rho_{jB}\frac{A_{ij}}{\sigma_{ij}^2} \hat \omega(\ell/\sigma_{ij}) 
 -\rho_W \frac{A_{iW}}{\sigma_{iW}^2} \hat \omega(\ell/\sigma_{iW}) \Big) \, ,
\end{split} \end{align}
where $\rho_{iA}$, $\rho_{iB}$ denote the number density of $i$th component in phases $A$ and $B$, $\rho_W$ is the density of the wall, and 
\beq
	\hat \omega(\ell) = 1-\ell \arctan \frac{1}{\ell} \, .
\eeq

For the following choice of the amplitudes and molecular sizes \cite{Dutka2008}
\begin{align} \begin{split}
& A_{ij}=\sqrt{A_{ii}A_{jj}} \, , \quad A_{iW}=\sqrt{A_{ii}A_{WW}} \, \\ 
& \sigma = \sigma_{ij} \, , \quad \sigma_W=\sigma_{iW}\, , \quad i,j=1,2
\end{split} \end{align}
the surface tension coefficient and the effective interface potential can be rewritten as
\begin{align} \begin{split}
  \gamma_{AB} =& \frac{\pi}{8 \sigma^2} \Big(\sum_{i=1}^2 \sqrt{A_{ii}}(\rho_{iB}-\rho_{iA})\Big)^2 \, \\
	\omega_{AB}(\ell)=& \frac{\pi}{4} \, \Big(\sum_{i=1}^2 \sqrt{A_{ii}}(\rho_{iB}-\rho_{iA})\Big) \, \\
	 & \quad \Big(\sum_{j=1}^2 \frac{\rho_{jB}\sqrt{A_{jj}}}{\sigma^2} \, \hat \omega(\ell/\sigma) -\frac{\rho_W\sqrt{A_{WW}}}{\sigma_W^2} \, \hat \omega(\ell/\sigma_W) \Big) \, .
\end{split} \end{align}
Upon introducing the dimensionless quantities
\begin{align} \begin{split}
\hat \rho_{AB} =& \frac{1}{2} \Big(1-\frac{\sum_{i=1}^2 \sqrt{A_{ii}}\rho_{iA}}{\sum_{i=1}^2 \sqrt{A_{ii}}\rho_{iB}} \Big) \,, \\
\hat A_{AB} = &\frac{\rho_W \sqrt{A_{WW}}}{\sum_{i=1}^2 \sqrt{A_{ii}}\rho_{iB}} \,, \\
 \hat \sigma_W =& \frac{\sigma_W}{\sigma}
\end{split} \end{align}
the effective interface potential reduces to 
\begin{align}
 \omega_{AB}(\ell) = \frac{\gamma_{AB}}{\hat \rho_{AB}}\Big[ \hat \omega_{AB} \Big(\frac{\ell}{\sigma}\Big) 
    - \frac{\hat A_{AB}}{\hat \sigma_W^2}\hat \omega_{AB} \Big(\frac{\ell}{\sigma}\frac{1}{\sigma_W} \Big) \Big] \, ,
\end{align}
which is exactly the form of the effective interface potential for the one component system, see Eqs\,(\ref{int2}) and \,(61) in Ref.\,\onlinecite{Dutka2012}. 

\begin{acknowledgments} F.D. was supported by Foundation for Polish Science within the project Homing Plus/2012-6/3, co-financed from European Regional Development Fund. M.N. acknowledges support from the National Science Center via grant 2011/03/B/ST3/02638.
\end{acknowledgments}

\bibliography{bibliography}

\end{document}